\documentstyle[12pt]{article}

\font\tenrm=cmr10

\newcommand{\bref}[1]{(\ref{#1})}
\newcommand{\ct}[1]{\cite{#1}}

\newcommand{\be}{\begin{equation}}
\newcommand{\ee}{\end{equation}}

\newcommand{\beq}{\begin{equation}}
\newcommand{\eeq}{\end{equation}}

\def\theequation{\thesection.\arabic{equation}}
\def\@eqnnum{{\rm (\theequation)}}

\def\lsim{\mathrel{\rlap{\lower4pt\hbox{\hskip1pt$\sim$}}
    \raise1pt\hbox{$<$}}}
\def\gsim{\mathrel{\rlap{\lower4pt\hbox{\hskip1pt$\sim$}}
    \raise1pt\hbox{$>$}}}
\def\frac#1#2{{{#1} \over{#2}}}

\begin{document}
\begin{titlepage}

\begin{flushright}
{SLAC-PUB-8747 \\ }
{LBNL-51153 \\ }
{\hfill April 2003 \\ }
\end{flushright}
\vglue 0.2cm

\begin{center}
{
{Comments on an Expanding Universe \\ }
\vglue 1.0cm
{Stuart Samuel$^{1}$ \\ }
\vglue 0.5cm

{\it Theory Group\\}
{\it Stanford Linear Accelerator Center\\}
{\it Stanford, CA 94309 USA\\}
\vglue 0.4cm
and
\vglue 0.4cm
{\it Theory Group, MS 50A-5101$^{2}$ \\}
{\it Lawrence Berkeley National Laboratory\\}
{\it One Cyclotron Road\\}
{\it Berkeley, CA 94720 USA\\}

\vglue 0.8cm


{\bf Abstract}
}
\end{center}
{\rightskip=3pc\leftskip=3pc
\quad Various results are obtained 
for a Friedmann-Robertson-Walker cosmology.  
We derive an exact equation that determines Hubble's law, 
clarify issues concerning the speeds of faraway objects and  
uncover a ``tail-light angle effect" 
for distant luminous sources. 
The latter leads to a small, previously unnoticed correction 
to the parallax distance formula. 
} 

\vfill          

\textwidth 6.5truein
\hrule width 5.cm
\vskip 0.3truecm
{\tenrm{
\noindent
$^{1)}$\hspace*{0.15cm}E-mail address: samuel@thsrv.lbl.gov \\ 
$^{2)}$\hspace*{0.15cm}Mailing address }}

\eject
\end{titlepage}

\newpage

\baselineskip=20pt

{\bf\large\noindent I.\ Introduction}\vglue 0.2cm
\setcounter{section}{1}
\setcounter{equation}{0}

The standard Friedmann-Robertson-Walker 
the metric\ct{frw} in spherical-like coordinates 
$r$, $\theta$ and $\phi$ is given by 
\beq
    c^2 d \tau^2 = 
    c^2 dt^2 - R^2(t)\left(  { { {dr^2} \over {(1 - k r^2)} } + 
     r^2 d \theta^2 + r^2 sin^2 \theta d \phi^2 } \right)
\label{frwmetric}
\quad . 
\eeq
where $R(t)$ is the cosmic scale factor 
at time $t$, 
$\tau$ is proper time,  
and $k$ determines whether three-space is
a sphere ($k=+1$), 
a flat space ($k=0$),
or
a hyperbolic sphere ($k=-1$).\ct{peebles,kolbturner,weinberg,bondi}

The metric in Eq.\bref{frwmetric} is a specific case of 
\beq
    c^2 d \tau^2 = 
    c^2 dt^2 - R^2(t)\left( g_{ij}(x) dx^i dx^j \right)
\label{genmetric}
\quad , 
\eeq
where $g_{ij}$ is a function of the spatial coordinates only. 
In going from time $t$ to a slightly later time $t^\prime $, 
each region of space stretches 
by the same factor $R( t^\prime  )/R(t)$. 
Due to this stretching, 
faraway objects are carried away from any particular observer 
moving with recessional speeds $v_r$ 
that increase with the distance:\ct{hubble} 
\beq
v_r = H r + \dots 
\label{hubblelaw}
\quad , 
\eeq
where $r$ is the distance to the object. 
The corrections to Eq.\bref{hubblelaw} vanish as $Hr/c \to 0$. 
Hence for $Hr/c \ll 1$, 
the linear Hubble law  
\beq
   v_{r} \approx H r
\label{linearhubblelaw}
\quad , 
\eeq 
is an excellent approximation.   
Hubble's constant is 
$
  H(t) = \dot R(t) / R(t) 
$.  

In obtaining Eq.\bref{linearhubblelaw}, 
it is assumed that the observer and the nearby objects 
do not have perculiar velocities. 
Throughout our work, 
we shall make use of a system of comoving observers. 
These comovers have values of the coordinates $x$ that are fixed in time 
so that, according to the metric in Eq.\bref{genmetric},  
the distance between two nearby comovers increases by 
a factor of $R(t_2)/R(t_1)$ between times $t_1$ and $t_2$. 

There are several commonly used distances 
to specify the spatial separation of a faraway object from Earth: 
the proper distance $d_{prop}$, 
the luminosity distance $d_{lum}$ 
determined by apparent brightness, 
the parallax distance $d_{parallax}$, 
the angular size distance $d_{as}$ obtained 
by measuring apparent width, 
and
the time-of-flight distance $d_{tof}$ given by 
$c (t_{obs} - t_{em})$ where $t_{em}$ is the time at which 
a distant object emits a light ray and $t_{obs}$ is the time 
at which it is observed on Earth. 
These distances agree 
with one another with increasing accuracy 
as the object approaches the Earth,
but differ significantly when the object is very faraway. 

Some of the above versions of distant violate the principle 
that instantaneous non-local measurements cannot be made. 
An example is proper distant. 
Speeds computed on the basis of proper distant are 
therefore  unphysical. 

Proper distant is an instantaneous  
measure of spatial separation 
that can be achieved only by engaging 
in a ``conspiracy" of multiple observers.  
Let $d_{prop} (t)$ be the proper distance 
between an object (e.g., a luminous source) 
and an observer (e.g., an astronomer). 
Arrange in advance for a series of comoving observers 
to be positioned between the two 
and instruct them to measure at the common time $t$ 
the distance to the next neighbor. 
See Figure 1. 
Let $\Delta x_{i+1,i}$ be the measured distance 
between observers $i$ and $i+1$. 
Then arrange for the observers to get together later 
to sum their measurements:  
\beq
   d_{prop} (t) = \sum_i \Delta x_{i+1,i} 
\quad . 
\eeq
Since, 
at a latter time $t'$, 
the distances $\Delta x_{i+1,i} $ all 
increase to $R(t')/R(t) \Delta x_{i+1,i} $,  
\beq
   d_{prop} (t') = {{ R(t') } \over { R(t) }} d_{prop} (t)  
\label{properdistancescaling}
\quad , 
\eeq
so that proper distance scales exactly 
with the cosmic scale factor. 

Define $v_{prop}(t)$ to be the rate of change of proper distance 
with respect to time: 
$v_{prop}(t) \equiv \partial d_{prop} (t) / \partial t$.
Then using $d_{prop}$ as {\it the} definition of distance 
and {\it assuming} $v_{prop}$ is the appropriate measure of speed, 
one would conclude that the Hubble law is exactly linear: 
\beq
   v_{prop}(t) = 
  {{ \partial  d_{prop} (t) } \over {  \partial t }} = 
  {{ \dot R (t) } \over { R(t) }} d_{prop} (t) = H(t) d_{prop} (t) 
\quad . 
\eeq 
Indeed, any definition of distance 
that scales exactly with $R(t)$ 
as in Eq.\bref{properdistancescaling} obeys 
such a linear Hubble law.
Since $d_{prop} (t) $ can be made arbitrarily large, 
one finds, 
with these definitions of speed and distance, 
that distant objects travel 
faster than the speed of light $c$.  
An interesting result in ref.\ct{stuckey} states that 
the proper distance for which $ H(t) d_{prop} (t) = c$ 
can actually occur within the particle horizon, 
that is, within the observable universe. 
It is sometimes stated that the Hubble law is exactly linear 
and that faraway objects can move away from Earth 
at a rate exceeding $c$.\ct{peebles,stuckey,murdoch} 
However, the use of $d_{prop}$ as {\it the} definition of distance 
is not physical as emphasized above 
in that it is impossible for an observer 
to make an instanteous measurement of it. 

However, for metrics of the form in Eq.\bref{genmetric}, 
there are definitions of distant 
that are physical and causal 
for which recessional speeds never exceed that of light. 
The basic idea is to use 
a dense network of comoving observers 
throughout the universe, 
who are allowed to make local measurements, 
that is, measurements in a small region centered 
about their positions. 
Non-local measurements are then achieved by 
communicating the local results to one another 
and by using relativistic dynamics.   
One needs to use relativistic dynamics because 
sizeable speeds enter for very distant objects.  
Any definition of distant that uses only local measurements 
and respects the principles of special relativity cannot 
lead to speeds of objects exceeding the speed of light. 

In the process of carrying out our analysis, 
we also uncover an angle effect 
not previously noted in general relativity.
The angle between two rays as measured by an observer 
in the vicinity of the rays but far from the source 
is not the same as the angle 
between the rays as emitted by the source. 
It is obvious that such an effect should exist: 
In special relativity, 
there is the ``tail light" effect: 
The light from a receding source is observed to spread out. 
Since distant objects 
are moving away from one another in an expanding universe, 
the ``tail light" effect should be present and, indeed, it is. 
This leads to a small correction to the standard formula 
for parallax distance. 

One way to illustrate how angles can change with time is as follows: 
Consider two nearby comoving observers. 
The two agree to send out light rays 
in a direction perpendicular to the line between them. 
See Figure 2. 
Then since space is expanding, 
the angle between the light rays 
will initially be slightly greater than zero 
and seen to increase with time by any local observer. 

The uncovering of the ``tail light" effect was actually 
the main motivation for the current work. 
Recent redshift data of distant type Ia supernovae suggest 
that the expansion of the universe is accelerating.\ct{sn}. 
This is contrary to what most cosmologists had expected. 
Physically, distant supernovae appear to be dimmer than expected. 
The ``tail light" effect could be the explanation 
if it has not been previously properly taken into account. 
However, using local comoving observer measurements, 
we obtain the standard formula 
for the luminosity distance of a light source. 
Hence, the ``tail light" effect is not the origin of
the unexpected faintness of distant supernovae. 
To account for an accelerating universe, 
a cosmological constant or 
some other dark energy contribution does need to be invoked. 

\medskip

{\bf\large\noindent II.\ An Exact (Differential) 
Hubble Law Equation}\vglue 0.2cm
\setcounter{section}{2}
\setcounter{equation}{0}

This section derives a new form of Hubble's law 
by determining the corrections to Eq.\bref{hubblelaw}. 
It is straightforward to obtain an exact equation 
for the recessional velocity $v$ 
as a function of distance $r$ from the Earth. 
First establish a network  
of comoving observers. 
Each observer sees the comovers in its vicinity 
moving away according to the Hubble law in Eq.\bref{linearhubblelaw}. 
See Figure 3. 

Suppose that the recessional speed $v(r)$ at $r$ 
has been determined using local measurements 
by comoving observers. 
Two speeds are involved in determining $v(r+\Delta r)$ 
at a slightly farther distance: 
(1) The comover at $r$ observes that a comover 
$\Delta r$ further out moves with a speed of $H \Delta r$ and 
(2) the comover at $r$ is moving away from Earth 
with a speed of $v(r)$. 
Using the relativistic formula for the addition of velocities, 
one finds that that the comover's speed at $r + \Delta r$ is 
$$
 v(r + \Delta r) = 
  {{ v(r) + H \Delta r } 
      \over { 1 + {{ v(r) H \Delta r } \over { c^2}} }}  
   \approx v(r) + H (1 - {{ v^2(r) } \over { c^2 }} ) \Delta r 
\quad , 
$$
or 
\beq
 { { d v(r) } \over { dr } } = H ( 1 - { { v^2(r) } \over { c^2 } } ) 
\label{exacthubblelaw} 
\quad . 
\eeq
Eq.\bref{exacthubblelaw} is a fundamental equation 
which can be integrated 
to obtain the exact recessional speed 
as a function of distance. 
For $r$ small, 
the $v^2(r) / c^2$ term can be neglected 
and one recovers the linear Hubble law. 

The formula for $v$ implicitly 
defines a distance $r$ by $dr/dt = v(t)$, 
which can be integrated out to determine $r$ if $r(t_0)$ is known 
for some early time $t_0$ and if the history of the universe 
is provided, that is, an exact formula for $H(t)$.  

If $H$ is constant, 
which is the case when the expansion 
is exponential ($R(t) = \exp (H t)$) 
and which might have happened 
in the early universe during inflation\ct{guth}, 
one finds 
\beq
  v(r) = c \tanh (Hr/c) = 
   c {{ \exp(Hr/c) - \exp (-Hr/c)  } 
    \over {  \exp(Hr/c) + \exp (-Hr/c) }} 
\label{exponentialexpansionspeed} 
\quad , 
\eeq
which yields a result for $v(r)$ 
that is always less than $c$ and approaches $c$ 
only for $r \to \infty$. 
It is sometimes misstated that,  
in an inflationary universe, 
superliminary speeds are achieved. 
For $r$ small, 
one recovers 
$v(r) = Hr + \dots$ from Eq.\bref{exponentialexpansionspeed}. 

In a Friedmann-Robertson-Walker universe, 
$H$ is not constant and one must integrate Eq.\bref{exacthubblelaw} 
taking into account the variation of $H$ with time. 
An example of how the integration is performed 
is provided below. 

Because only local measurements are made that respect 
the principles of special relativity 
in deriving Eq.\bref{exacthubblelaw}, 
no object is viewed as having 
a recessional speed greater that $c$. 
Indeed, 
as $v(r)$ approaches $c$, 
the factor $(1 - v^2(r) / c^2)$ in Eq.\bref{exacthubblelaw} 
reduces the incremental increase in speed. 
In support of Eq.\bref{exacthubblelaw}, 
we have been able to show that 
a number of results pertinent to an expanding universe 
are obtainable from the differential Hubble law. 
Here, we restrict ourselves to only one example:
a derivation of the redshift 
as a Doppler effect.  

The expansion of the universe 
causes the light from a distance source to be shifted to the red
because the wavelength $\lambda$ of light 
is stretched:   
\beq
   \lambda (t_{obs}) = 
   {{ R( t_{obs} ) } \over {  R( t_{em} ) }}  \lambda (t_{em}) 
\label{redshift}
\quad , 
\eeq
where $\lambda (t_{obs}) $ is the wavelength of the light 
at the time that it is observed and 
$\lambda (t_{em}) $ is the wavelength at the time that it is emitted. 
Eq.\bref{redshift} holds for any metric of the form in
\bref{genmetric}. 

Let us show how to obtain 
the redshift as a Doppler effect due to the recessional velocity. 
If a luminous source 
is receding from an observer at a speed $v_{obs}$ then 
in special relativity 
\beq
  \lambda (t_{obs}) = 
    \sqrt{ {{ c + v_{obs} } \over { c - v_{obs} }} } \lambda (t_{em}) 
\label{dopplerformula}
\quad . 
\eeq
To determine $v_{obs}$, 
one integrates Eq.\bref{exacthubblelaw} 
for the process in which light 
is emitted from a distant object 
and received by an observer on Earth. 
As the light propagates, 
the change in distance $dr$ that it travels  
in $dt$ is given by $dr = - c dt$. 
Using this in Eq.\bref{exacthubblelaw} gives 
\beq
   dv = - \left( { 1 - {{ v^2 } \over { c^2 }} } \right) c H(t) dt 
\label{dv}
\quad . 
\eeq 
Separating variables and  
integrating from the initial time $t_{em}$ 
to the final time $t_{obs}$,  
one obtains  
\beq
  \sqrt{ {{ c + v_{obs} } \over { c - v_{obs} }} } = 
   \exp (\int_{t_{em}}^{t_{obs}} H(t) dt )   
   = { { R( t_{obs} ) } \over  { R( t_{em} ) } } 
\label{velocityscalerelation} 
\quad ,  
\eeq 
where the last equality holds because $H = d log(R)/ dt$.
Substituting this result into Eq.\bref{dopplerformula} 
yields the result in Eq.\bref{redshift}. 
The derivation supports the validity 
of the differential Hubble law 
in Eq.\bref{exacthubblelaw} and 
illustrates how it is necessary to take into consideration 
the variation of the Hubble constant  
in integrating the equation. 

It is sometimes stated that the redshift cannot 
be computed as a Doppler effect. 
The argument goes as follows. 
Suppose that one can vary the expansion factor $R(t)$ at will. 
Around the time of emission, 
adjust $R(t)$ so that it is constant. 
After emission, 
let $R(t)$ increase so that the universe expands 
and produces a redshift. 
Before observation, 
adjust $R(t)$ so that it is constant again. 
Then one might argue that, 
since the universe 
is not expanding during emission and observation, 
there is no relative velocity between emitter and observer 
during these processes, 
and hence no Doppler effect. 
There are several difficulties with this reasoning. 
First, 
it assumes that the relative speed between two distant objects 
can be instantaneously measured and 
hence is zero at the times of emission and observation. 
Second, 
the above derivation leading 
to Eq.\bref{velocityscalerelation} 
demonstrates unequivocally that the red shift {\it can} 
be computed as a Doppler effect for arbitrarily varying $R(t)$. 
It is clear from this computation that the recessional speed 
is ``built up" during the entire period of light propagation 
and is not instantaneously produced. 
Third,  
changing $R(t)$ from a constant to a non-zero value  
creates an acceleration between the light and the observer 
(and also with the emitter).  
This acceleration generates a redshift 
as can can see as follows. 
Consider the line of comovers positioned 
between the source and final observer on Earth. 
Let each of these intermediate comovers 
absorb the light and instantly re-emit it. 
This has no effect on the light. 
During the periods for which $R(t)$ is unchanging, 
no redshift is generated. 
However, as soon as $R(t)$ increases, 
two nearby intermediate comovers achieve a relative velocity 
and the next one observes a redshift compared to the previous one 
that can be attributed to the acceleration of space 
or as a Doppler effect. 

It is incorrect to incorporate both the Doppler effect 
and the stretching of space in determining 
$ {{ \lambda ( t_{obs} ) } / {  \lambda ( t_{em} ) }}$: 
The redshift in general relativity in an expanding universe 
is due to the stretching of waves of light; 
Observers, however, have the option of viewing the 
the redshift as due a Doppler effect 
arising from  the relative motion  
of sources and observers. 

If a distant source has a peculiar velocity, 
then in general relativity there is both a cosmological redshift 
and a moving-source Doppler effect. 
It is easily checked that 
the use of \bref{dv} correctly produces
the total wavelength shift as a single Doppler effect 
by using $v_{peculiar}$ for the initial speed 
and noting that the velocity addition formula in  
special relativity $v_3 = (v_1 + v_2)/(1 + v_1 v_2 /c^2)$ 
can be written as 
\beq
   \sqrt{ {{ c + v_{3} } \over { c - v_{3} }} } = 
   \sqrt{ {{ c + v_{1} } \over { c - v_{1} }} }
   \sqrt{ {{ c + v_{2} } \over { c - v_{2} }} }
\label{velocityaddition} 
\quad , 
\eeq 
as can easily be checked. 

\medskip

{\bf\large\noindent III.\ The ``Tail-Light'' Effect }\vglue 0.2cm
\setcounter{section}{3}
\setcounter{equation}{0}

In this section, 
we compute the luminosity distance 
using a network of comoving observers 
because, among things, it allows us to  
uncover a ``tail-light angle effect." 
In addition, 
since the luminosity distance is used 
in analyzing type Ia supernova data, 
a careful, detailed derivation of the formula 
is worth performing 
given that the tail-light effect has previously been overlooked. 

If the absolute luminosity of an astrophysical object is known, 
then its apparent luminosity $L_A$ can be used to determine 
a distance $d_{lum}$ to the object\ct{weinberg}: 
\beq
  L_A \propto  {  { R^2 (t_{em}) } 
  \over { R^2 (t_{obs}) } }  { {1} \over {d_{eff}^2} } 
\label{apparentluminosity}
\quad , 
\eeq
where  
$d_{eff}$, the effective distance, is defined as 
\beq
   b_{obs} \equiv d_{eff} \phi_{s} 
\quad , 
\eeq
and $b_{obs}$ is the impact parameter 
or distance measured by the observer of two nearby light rays 
emitted from the source 
with an angular separation at the source of $\phi_{s}$. 
See Figure 4. 
The definition of luminosity distance $d_{lum}$ is   
\beq
  d_{lum} \equiv {{ R( t_{obs} ) } \over {  R( t_{em} ) }}  d_{eff} 
\label{dlumdef} 
\quad . 
\eeq 

It remains to determine $d_{eff}$. 
Arrange for a set of equally spaced, 
intermediate comoving observers 
to be between the source and the receiver. 
Because the angle $\phi_s$ is typically extremely small, 
terms of order $\phi_s^2$ may be neglected. 

Figure 5(a) shows the initial emission of the two rays. 
Angles are display much larger than the actual case for clarity. 
Not surprisingly, 
observers disagree on what transpires 
as the rays move from the source 
to the intermediate comoving observer 1. 
According to observer 1, 
the light is emitted at a distance $\Delta x$ at an angle $\phi_1$, 
and it travels to the region of 1 
while the source moves away at a recessional speed of $v_{1s}$. 
As the rays pass 1 at time $t_1$, 
the distance between the source and 1 becomes 
$R(t_{1}) / R(t_{em}) \Delta x$ 
because space is expanding. 
The angle $\phi_1$ is greater than $\phi_s$ 
because for a source that is moving away, 
one has the ``tail light" effect of special relativity. 
See Figure 5(b). 
Using the standard formula for the relation 
between angles in special relativity 
for references frames moving with respect to one another, 
\beq
   \phi_1 = 
   \sqrt{ {{ c + v_{1s} } \over { c - v_{1s} }} }  \phi_s = 
      {{ R(t_{1}) } \over { R(t_{em}) }} \phi_s 
\label{anglerelation}
\quad , 
\eeq
where the last equality 
follows from Eq.\bref{velocityscalerelation}. 
The distance between the two rays $b_1$ 
as the light passes 1 is, according to observer 1, 
\beq
  b_1 = \Delta x \phi_1 = 
  {{ R(t_{1}) } \over { R(t_{em}) }} \Delta x \phi_s  
\label{b1}
\quad . 
\eeq 

According to the comoving observer at the source, 
the rays are emitted with an angular separation of $\phi_s$, 
but, as the rays move from the source to 1, 
observer 1 moves away. 
The distance the light travels is $R(t_{1}) / R(t_{em}) \Delta x $ 
and thus greater. 
Using this distance and angle, 
an observer at the source arrives at Eq.\bref{b1} 
for the impact parameter at 1, 
in accord with special relativity 
that moving observers agree on distances perpendicular 
to their relative motion. 
See Figure 5(c). 

Now consider the process 
in which the rays travel from the region 
of intermediate comover 1 
to the region of intermediate comover 2. 
It is convenient to consider a comoving observer at 1$^\prime$ 
where the upper ray passes near 1 and 
a corresponding comoving observer at 2$^\prime$ as in Figure 6(a). 
Because the observer at 1$^\prime$ is moving away from 1, 
the angle that the upper ray makes 
with a horizontal line perpendicular to the line running 
between 1 and 1$^\prime$ is less than $\phi_1$. 
In fact, 1$^\prime$ is moving away with just the right speed 
to observe the angle as $\phi_s$ to order $\Delta x ^2$. 
The situation for 1$^\prime$ and 2$^\prime$ in Figure 6 
is thus similar to that of 1 and the source in Figure 5 
except that the distance between 1$^\prime$ and 2$^\prime$ 
is $R(t_{1}) / R(t_{em}) \Delta x $ instead of $\Delta x$. 
Let $\Delta b_2$ be the impact parameter seen by 2$^\prime$. 
Then 
\beq
  \Delta b_2 = 
    {{ R(t_{2}) } \over { R(t_{1}) }} 
      {{  R(t_{1}) } \over { R(t_{em}) }}  \Delta x \phi_s 
        =  {{  R(t_{2}) } \over { R(t_{em}) }} \Delta x \phi_s  
\quad . 
\eeq 
Since the primed comoving observers 1$^\prime$ and 2$^\prime$  
are moving away from the unprimed observers 1 and 2 
due to the expansion of the universe, 
the distance between them increases 
as the rays move from 1 to 2. 
Hence, the distance between primed and unprimed observers 
is $R(t_2)/R(t_1) b_1$ when the rays arrive in region 2 
and the impact parameter $b_2$ at 2 is 
\beq
  b_2 = 
   {{ R(t_2) } \over { R(t_1) }} b_1 +  
   {{ R(t_{2}) } \over { R(t_{em}) }}  \Delta x \phi_s  = 
  {{ R(t_{2}) } \over { R(t_{em}) }} 2 \Delta x \phi_s  
\label{b2}
\quad . 
\eeq
As in the case of Figure 5, 1$^\prime$ and 2$^\prime$ disagree 
on what happens as the rays move from region 1 to 2 
but agree on the value of $b_2$. 
See Figures 6(b) and 6(c). 

The process in which rays go from comoving observer $i$ to $i+1$ 
is similar to that of Figure 6 except the distance 
between $i$ to $i+1$ is initially $R(t_i)/R(t_{em}) \Delta x$ 
and the angle for the upper ray at $i'$ is larger 
and equal to $\phi_i$ as seen by $i$. 
The comoving observer at $i'$, however, 
sees the angle as $\phi_s$. 
In place of Eq.\bref{b2}, one has 
\beq
  b_{i+1} = 
   {{ R(t_{i+1}) } \over { R(t_i) }} b_i +  
      {{ R(t_{i+1}) } \over { R(t_{em}) }}  \Delta x \phi_s  =  
   {{ R(t_{i+1}) } \over { R(t_{em}) }} (i+1) \Delta x \phi_s  
\label{bip1}
\quad . 
\eeq

Equation \bref{bip1} can be evaluated 
at the position of the receiver 
by setting $i=N-1$ for $N-1$ intermediate observers:
\beq 
 b_{obs} =  
 {{ R(t_{obs}) } \over { R(t_{em}) }} \sum_i  \Delta x \phi_s  
\quad . 
\label{bobs}
\eeq
Since $\sum_i \Delta x = N \Delta x = d_{prop}(t_{em})$, 
one concludes that
\beq 
  d_{eff} = 
   {{ R(t_{obs}) } \over { R(t_{em}) }} d_{prop}(t_{em}) = 
     d_{prop}(t_{obs}) 
\label{deff}
\quad , 
\eeq
so that the luminosity distance in Eq.\bref{dlumdef} is 
\beq
 d_{lum} = {{ R(t_{obs}) } \over { R(t_{em}) }} d_{prop}(t_{obs})  
\label{dlum}
\quad , 
\eeq
which agrees with the standard result.\ct{peebles,weinberg,robertson} 
Although there is a ``tail light" effect, 
it is not the reason why distant type Ia 
supernovae appear dimmer than expected.

The angle $\phi_i$ measured by the $i$th observer 
at time $t_i$ is 
\beq 
  \phi_i \approx 
   \left( \phi_s + 
        {{ H(t_i) b_i } \over { c }} \right) 
   = \phi_s \left( 1 + { H(t_i) R(t_i) } \int_{t_{em}}^{t_i}  
       {{ds} \over {R(s)}}  \right) 
\label{angleofray} 
\quad . 
\eeq
This is the ``tail light" effect: 
$\phi_i > \phi_{i-1} > \phi_s$. 
Since the ``tail light" effect can be quite significant 
for very distant astronomical luminous objects, 
one might wonder why it has not been detected experimentally. 
The reason is that, 
although $\phi_{obs}$ can differ by $\phi_s$ by a sizeable factor, 
both $\phi_{obs}$ and $\phi_s$ are miniscule 
and hence not directly measurable in practice. 
For example, suppose the mirror of a telescope 
is $1$ meter so that $b_{obs} \sim 1$ meter 
and that a supernova is observed with a redshift of $z \approx 0.5$. 
Then $\phi_{obs} / \phi_s \approx 1.5$ 
but the order of magnitude of either angle is $10^{-26}$ radians. 

It only takes a ``small factor" 
within the framework of an $\Omega = 1$  
Friedmann-Robertson-Walker universe 
to obtain agreement 
with the type Ia supernova observations. 
If the $i$th primed observer 
would have observed 
the angle of the upper ray as $R(t_i)/R(t_{i-1})$ 
times the angle observed by the $(i-1)$th primed observer 
(instead of an angle of $\phi_s$), 
then one would have found a somewhat larger value 
of $d_{eff}$ of $ {{ R(t_{obs}) } \over { R(t_{em}) }} d_{tof}$ 
and $d_{lum} $ would become 
$ {{ R^2(t_{obs}) } \over { R^2(t_{em}) }}  d_{tof}$, 
which turns out to fit the supernova data\ct{sn} perfectly 
for a flat-space universe. 
More specifically, 
in a $k=0$ matter-dominated universe, 
$d_{eff} = d_{prop} = 2 c (1 - 1/\sqrt{1+z}) /H_0 $, 
whereas 
$d_{eff}^{phen} = 
{{ R^2(t_{obs}) } \over { R^2(t_{em}) }}  d_{tof} = 
3 c (1 + z - 1/\sqrt{1+z} )/(2 H_0)$. 
Since the use of $d_{eff}^{phen}$ fits the supernova data  
so well, 
it can be used as a phenomenological parametrization
in current and future studes of the acceleration of the universe.

\medskip

{\bf\large\noindent IV.\ A Correction to the 
Formula for Parallax Distance}\vglue 0.2cm
\setcounter{section}{4}
\setcounter{equation}{0}

The definition of parallax distance is 
\beq
 d_{parallax} = {{ b_{obs} } \over {\phi_{obs} }}
\quad , 
\label{parallaxddef}
\eeq 
where $\phi_{obs} $ and $b_{obs}$  
are respectively the observed angle and distance between two rays. 
See Figure 4. 
Because of the ``tail light" effect, $\phi_{obs} $ 
is greater than $\phi_s$, 
and the location of the source appears closer 
than otherwise would be the case. 

Using the results for $b_{obs} $ and $\phi_{obs}$  
in Eqs.\bref{bobs} and \bref{angleofray} of the last subsection, 
one finds 
\beq
 d_{parallax} = 
  {{ d_{prop} (t_{obs}) 
   (1 - k d_{prop} (t_{obs})^2 / R(t_{obs})^2 )^{-1/2} } 
    \over { 1 + H(t_{obs} ) d_{prop} (t_{obs}) /c   }}
\quad , 
\label{correctparallax}
\eeq
which differs from the standard result\ct{weinberg} 
by the factor in the denominator. 
It is important to note that since parallax measurements in astronomy 
are made at positions fixed to the center of the solar system, 
non-comoving observers are used. 
This is the reason why $\phi_{obs}$ 
should be used and not $\phi_s$ 
(compare unprimed and primed observers of the last subsection). 
Since parallax is currently only used 
for relatively nearby astrophysical objects, 
the denominator correction factor numerically does not 
significantly affect parallax distant measurements.

\medskip

{\bf\large\noindent V.\ Conclusions}\vglue 0.2cm
\setcounter{section}{5}
\setcounter{equation}{0}

In this research, 
we clarified several issues concerning the physics 
of a Friedmann-Robertson-Walker cosmology 
and derived several new results. 
In particular, with the use of reasonable definitions, 
recessional speeds no longer exceed the speed of light. 
More importantly, 
we obtained a new, 
fundamental equation governing Hubble's law. 
There are statements in the literature that recessional speeds 
can exceed $c$ and that the Hubble law is exactly linear, 
but they are based on definitions requiring 
non-local instantaneous measurements.
We found
a correction factor for parallax distance 
that had previously been overlooked.  
Another new result is that 
the light rays from a distant source spread out. 
This ``tail light" effect, however, 
does not explain why recent distant type Ia supernovae 
appear dimmer than expected and 
therefore does not provide a way of avoiding 
the conclusion that the supernova data 
supports an accelerating expanding universe. 
Finally, 
we have uncovered a nice phenomenological fit 
for the type Ia supernova data. 

\medskip

{\bf\large\noindent Acknowledgments}

This work was supported in part
by the National Science Foundation under the grants 
(PHY-9420615 and PHY-0098840) and the Department of Energy 
under contract numbers DE-AC03-76SF00515 and 
DE-AC03-76SF00098.
I thank V.\,Parameswaran Nair, Bob Wagoner 
and Michael Peskin for discussions. 
I also thank Marty Tiersten for positive feedback 
and W.\,M.\,Stuckey for pointing out some typographical errors 
in the original manuscript. 

\bigskip

\def\PRL#1#2#3{ {Phys.{\,}Rev.{\,}Lett.{\,}}{\bf {#1}}, {#2} ({#3})}
\def\PRD#1#2#3{ {Phys.{\,}Rev.{\,}}{\bf D{#1}}, {#2} ({#3})}
\def\PR#1#2#3{ {Phys.{\,}Rep.{\,}}{\bf {#1}}, {#2} ({#3})}
\def\OPR#1#2#3{ {Phys.{\,}Rev.{\,}}{\bf {#1}}, {#2} ({#3})}

\pagebreak
{\bf\large\noindent Figure Captions}  

\noindent
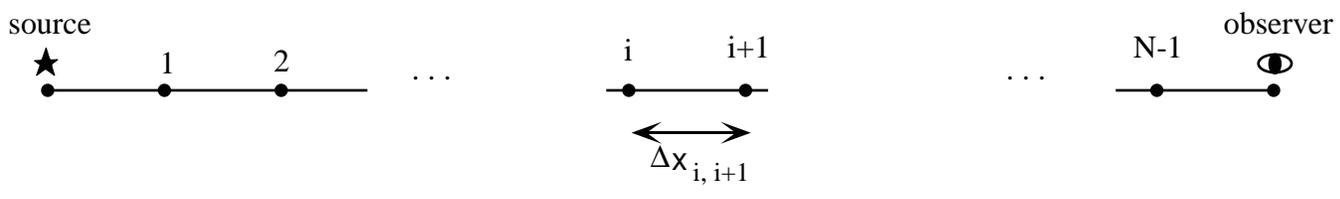
Figure 1. Arrangement of Intermediate Comoving Observers 
to Compute the Proper Distance Between a Source and an Observer 

\noindent
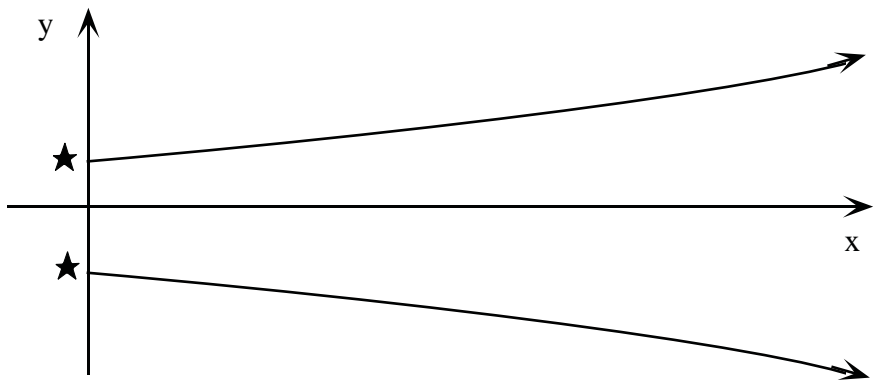
Figure 2. The Spreading Out of Rays Emitted 
in the x-Direction by Separated Comoving Sources

\noindent
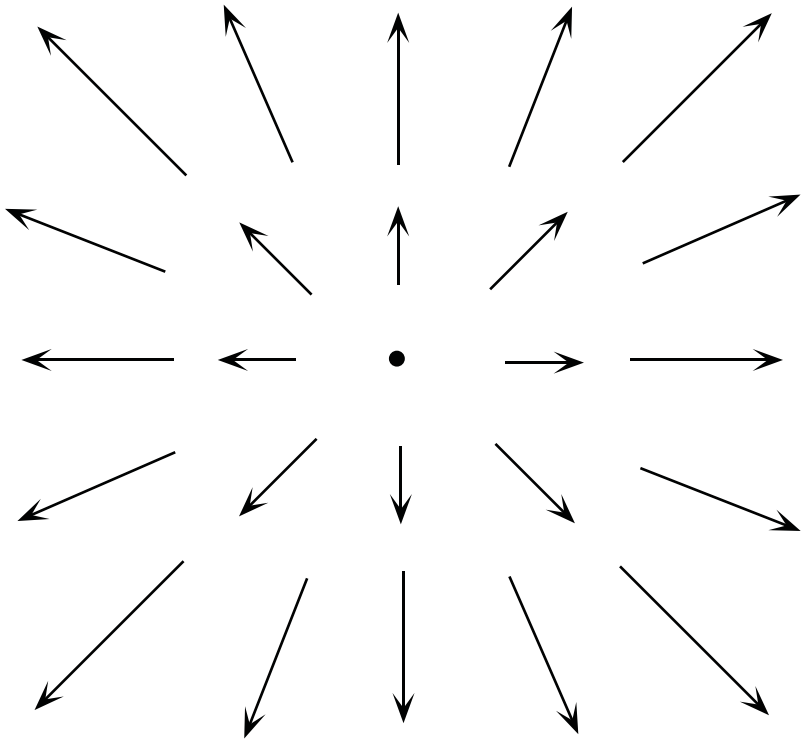
Figure 3. The Recessional Velocity Vectors in 
a Region of a Comoving Observer

\noindent
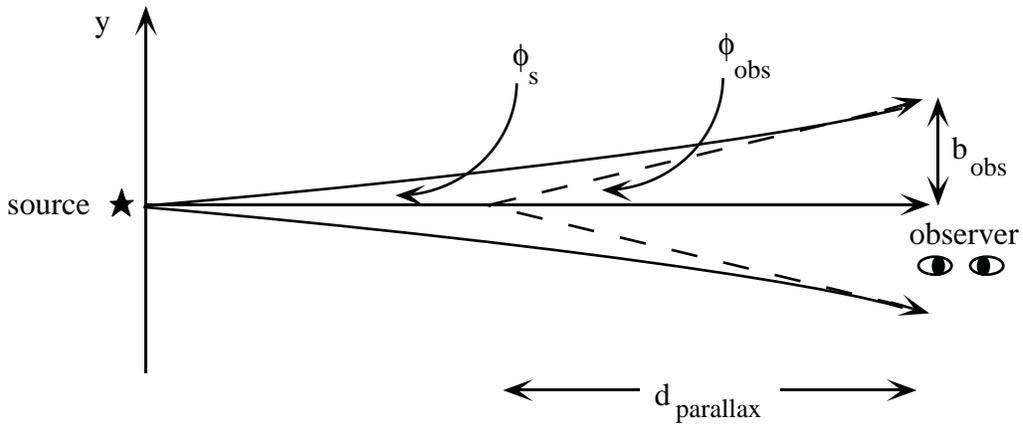
Figure 4. Rays Emitted at Small Angles 
by a Source and Measured by a Faraway Observer \\
The angle between the rays seen by the observer 
is larger than the angle at emission 
so that when the rays are projected back, 
they converge on a distance $d_{parallax}$, 
which is closer than one would obtain 
if the ``tail light" angular effect were to be neglected. 

\noindent
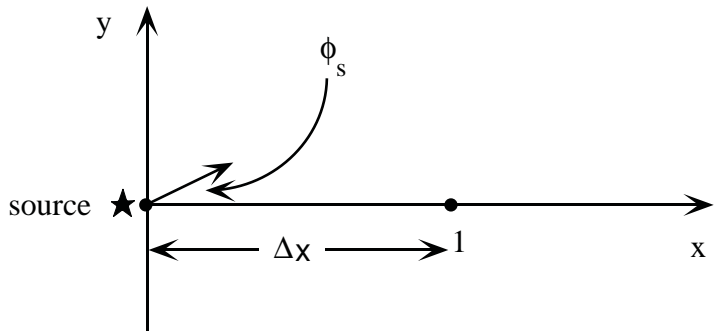
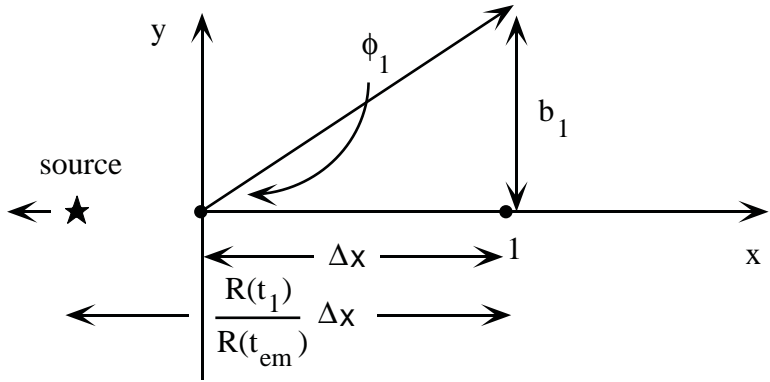
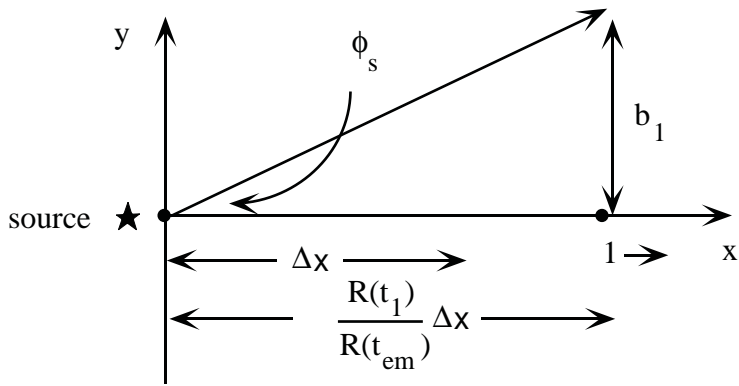
Figure 5. The Computation of the Impact Parameter $b_1$ 
at the First Intermediate Observer; 
(a) The Initial Situation at Time $t_{em}$ 
as Viewed by Observer 1; 
(b) The Process from the Viewpoint of Observer 1; 
(c) The Process from the Viewpoint of an Observer at the Source. 

\noindent
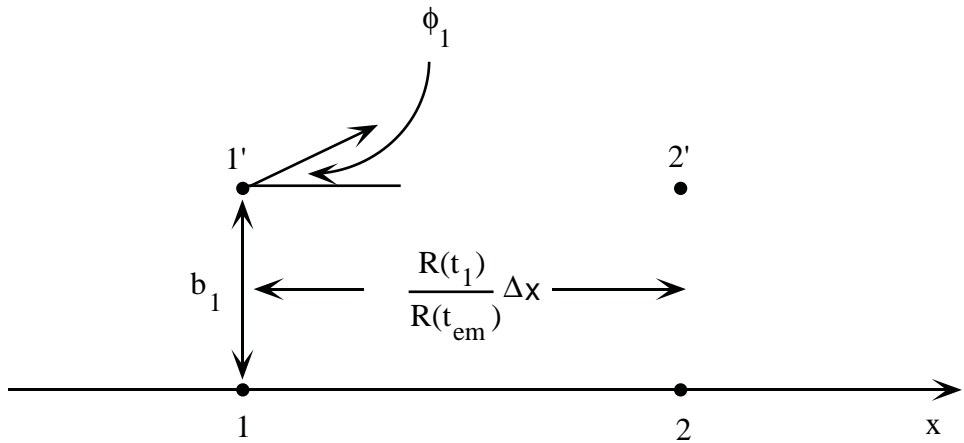
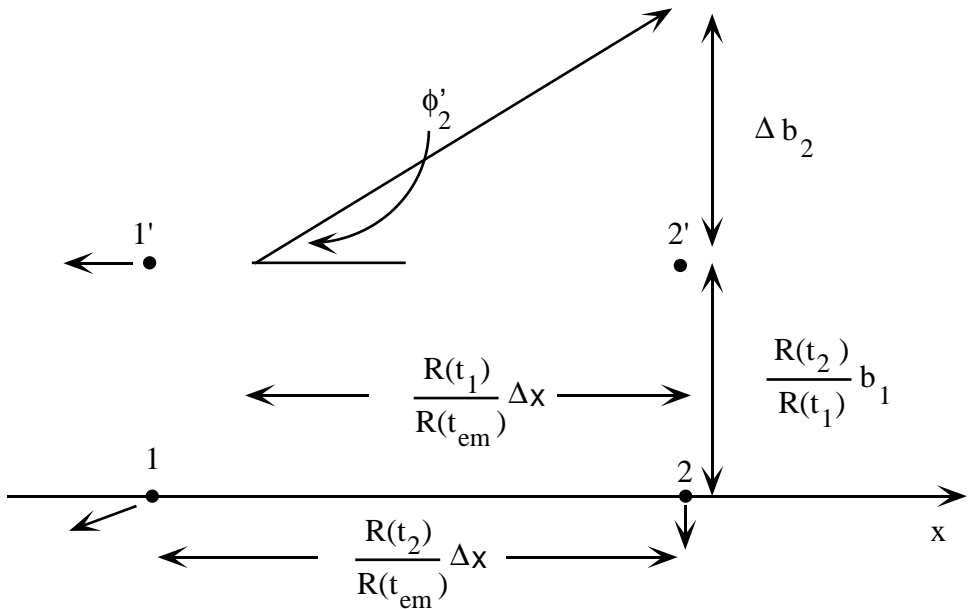
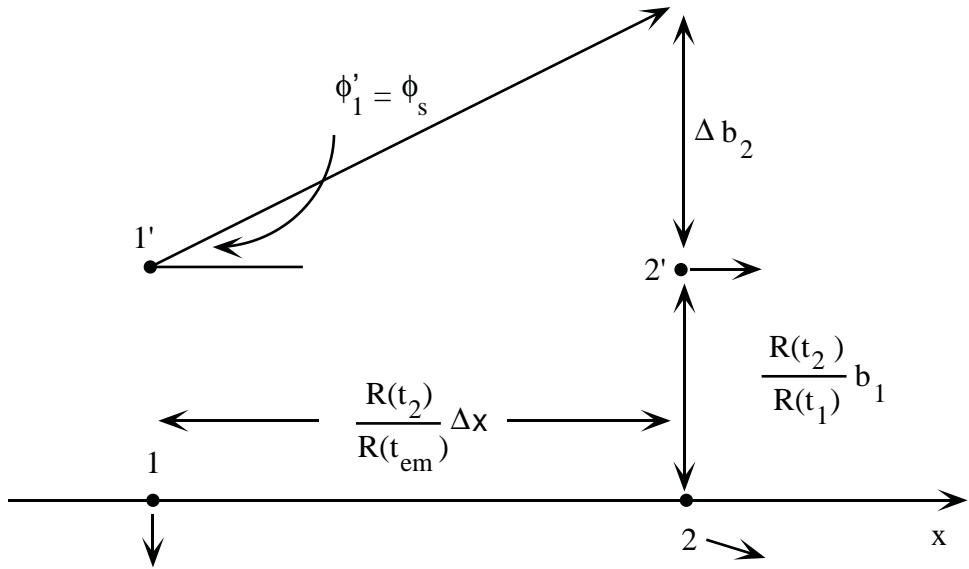
Figure 6. The Computation of the Impact Parameter 
at the Second Intermediate Observer; 
(a) The Situation When the Ray Passes 1 at Time $t_1$; 
(b) The Process from the Viewpoint of Observer 2$^\prime$; 
(c) The Process from the Viewpoint of Observer 1$^\prime$.

\vfill\eject
\end{document}